\documentclass[aps,prb,twocolumn]{revtex4-1}
\usepackage{graphicx}
\usepackage{amssymb}
\usepackage{amsfonts}
\def\checkT{{\check{T^{\vphantom{'}}}} }

\newcommand{\be}{\begin{equation}}
\newcommand{\bd}{\begin{displaymath}}
\newcommand{\ee}{\end{equation}}
\newcommand{\ed}{\end{displaymath}}
\newcommand{\ba}{\begin{eqnarray}}

\newcommand{\ea}{\end{eqnarray}}

\newcommand{\halfs}{{\scriptstyle \frac{1}{2}}}
\newfont{\mycal}{eufb10 at 12pt}
\newfont{\myeu}{eufm10}

\newcommand{\e}{{\rm e}}

\begin{document}
\title[Layered Ising Model]%
{Criticality in Alternating Layered Ising Models : \\ I.
Effects of connectivity and proximity}

\author{Helen Au-Yang}
\affiliation{Department of Physics, Oklahoma State University,
145 Physical Sciences, Stillwater, OK 74078-3072, USA}
\email{helenperk@yahoo.com}
\author{Michael E Fisher}
\affiliation{Institute for Physical Science and Technology,
University of Maryland, College Park, MD 20742-8510}
\email{xpectnil@umd.edu}

\begin{abstract}
The specific heats of exactly solvable alternating layered planar
Ising models with strips of width $m_1$ lattice spacings and
``strong'' couplings $J_1$ sandwiched between strips of width
$m_2$ and ``weak'' coupling $J_2$, have been studied numerically
to investigate the effects of connectivity and proximity. We find
that the enhancements of the specific heats of the strong layers
and of the overall or `bulk' critical temperature,
$T_c(J_1,J_2;m_1,m_2)$, arising from the collective effects
reflect the observations of Gasparini and coworkers in experiments
on confined superfluid helium. Explicitly, we demonstrate that
finite-size scaling holds in the vicinity of the upper limiting
critical point $T_{1c}$ ($\propto J_1/k_B$) and close to the
corresponding lower critical limit $T_{2c}$ ($\propto J_2/k_B$)
when $m_1$ and $m_2$ increase. However, the residual
{\it enhancement}, defined via appropriate subtractions of
leading contributions from the total specific heat, is dominated
(away from $T_{1c}$ and $T_{2c}$) by a decay factor $1/(m_1+m_2)$
arising from the {\it seams} (or boundaries) separating the strips;
close to $T_{1c}$ and $T_{2c}$ the decay is slower by a factor
$\ln m_1$ and $\ln m_2$, respectively.
\end{abstract}
\maketitle
\section {Introduction}

Many experiments performed on $^4$He at the superfluid transition
in various spatial dimensions,\cite{GKMD} reveal excellent agreement
with general finite-size scaling theory.\cite{Fisher, Barber}
Furthermore, when small boxes or ``quantum dots'' of helium were
coupled through a thin helium film, effects of connectivity and
proximity were discovered and
quantified.\cite{PKMG,KMG,PKMGn,MEF,PG,PKMGpr} 

\begin{figure}[!hbtp]
\centering
\includegraphics[width=\hsize]{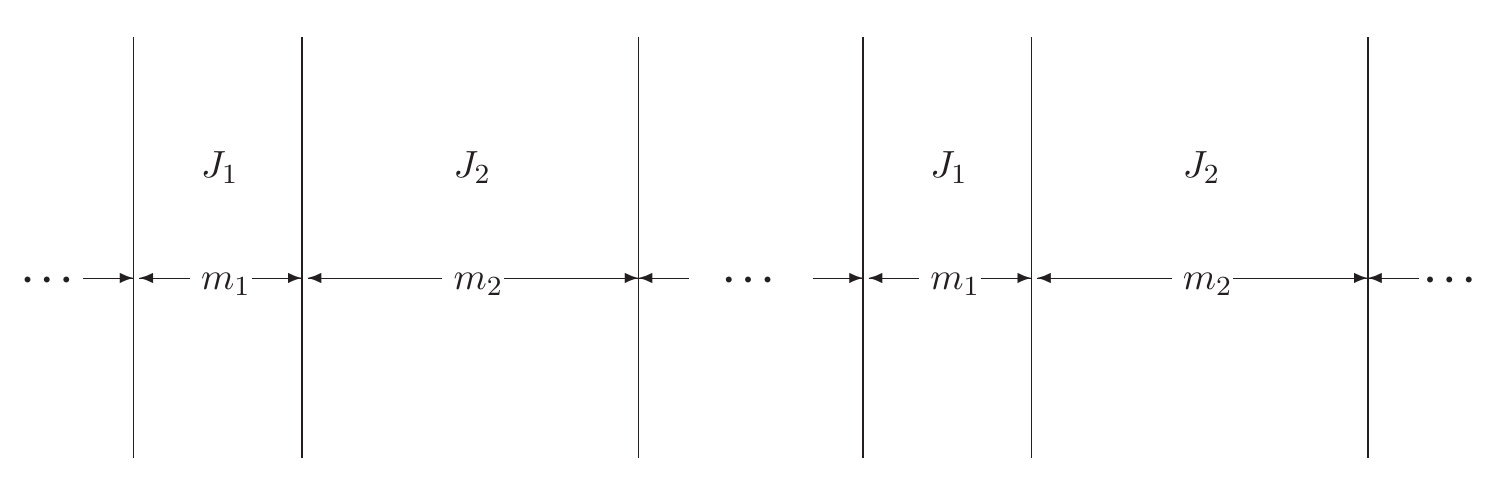}
\caption{The planar, square lattice alternating layered Ising
model considered. The widths $m_1$ and $m_2$ are measured in
nearest-neighbor lattice spacings, $a$, while nearest-neighbor
Ising spins $\sigma_i=\pm1$ are coupled via pair Hamiltonians
$J_{ij}\sigma_i\sigma_j$ with $J_{i,j}=J_1$ or $J_2$ as illustrated
schematically. On the seams at lattice sites with $x=n(m_1+m_2)a+a$
for $n=0,\pm1,\pm2,\dots$, the vertical bonds are of energy $J_1$
while the horizontal bonds are of strength $J_1$ on the right but
$J_2$ on the left; conversely, for the seams at
$x=(n+1)m_1a+nm_2a+a$  the vertical bonds have strength $J_2$
while the horizontal bonds on the right are of energy $J_2$
but $J_1$ on the left.}
\label{fig1}
\end{figure}

To gain some more detailed theoretical insights into the proximity
effects, we study here the specific heats of an alternating layered
planar Ising model, which consists of infinite strips of width
$m_1$ lattice spacings in which the coupling or bond energy between
the nearest-neighbor Ising spins is $J_1$, separated by other
infinite strips of width $m_2$ bonds (or lattice spacings) whose
coupling $J_2$ is weaker. This is illustrated in Fig.~\ref{fig1}.

\begin{figure}[htbp]
\centering
\includegraphics[width=0.97\hsize]{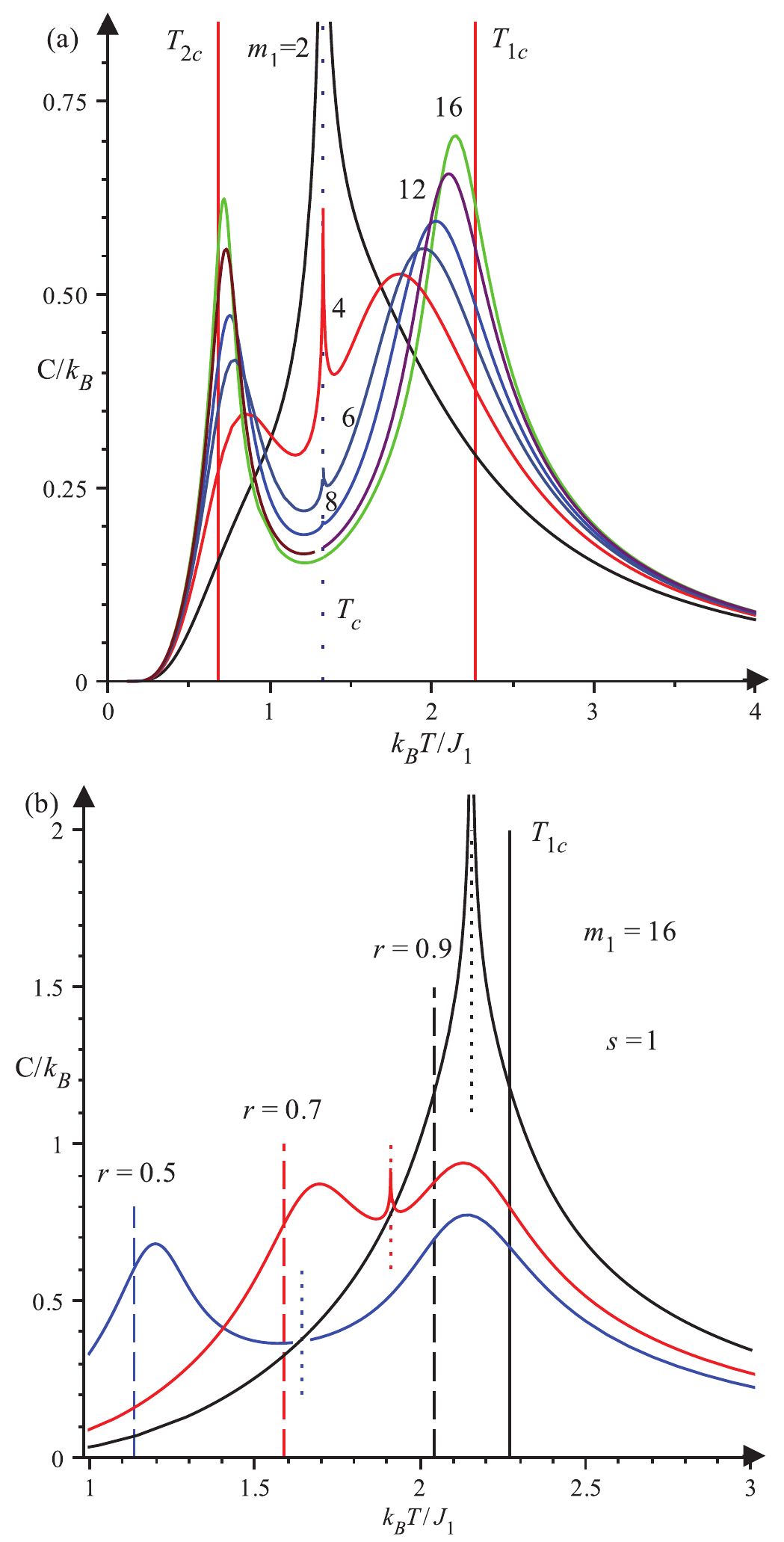}
\caption{(Color online) (a) The specific heats per site with
relative strength $r=J_2/J_1=0.3$ and relative separation
$s=m_2/m_1=1$ for $m_1=2,4,\cdots,16$. The amplitude $A(r,s)$ of
the logarithmic divergence at $T_c(r,s)$, which dominates for
$m_1=2$, decreases rapidly as $m_1$ increases. Thus, the small
spike at the ``true'' bulk critical point, $T_c$ (indicated by
the dotted vertical line), becomes barely visible for $m_1\ge8$.
\break
(b) Plots of the specific heats for fixed $m_1=m_2=16$ and so
$s=1$ but for increasing relative strength $r$. Note that the
logarithmic peak at the overall or bulk critical point, $T_c(r,s)$,
indicated by short vertical dotted lines, remains clearly visible
when $r=0.7$ and still dominates entirely when $r=0.9$. The
vertical  dashed lines denote the positions of $T_{2c}(r)$. Unlike
part (a), now the spike remains evident at $m_1=16$ when
$r\uparrow1$. However, as $A(r,s)$ becomes small, two quite
distinct rounded peaks appear moving toward the limiting values
$T_{2c}(r)$ (denoted by vertical dashed lines) and  $T_{1c}$,
as $m_1=m_2\to\infty$.}
\label{fig2}
\end{figure}

When $J_2$ vanishes, the model becomes a system of noninteracting
infinite strips of finite width, each of which essentially behaves
as a one-dimensional Ising model. This means, in particular, that
the specific heat is not divergent but rather has a fully analytic
rounded peak. However, as long as $J_2\ne0$, the system is a
two-dimensional bulk Ising model, whose specific heat per site
diverges logarithmically at a unique bulk critical temperature
$T_c(J_1,J_2;m_1,m_2)$ in the form
\be
C(T)/k_B\propto-A(r,s)\ln|1-(T/T_c)|+B(r,s)+\cdots,\label{log}
\ee
where we have introduced the basic weakness or coupling ratio, $r$,
and the relative separation distance, $s$, namely
\be
\quad r=J_2/J_1<1,\quad s=m_2/m_1.
\label{dfrs}\ee

In fact, as will be shown in Part II,\cite{HAY} the amplitude
$A(r,s)$ of the logarithmic divergence decays exponentially as a
function of $m_1$ or of $m_2$;\cite{HAY}  indeed, at fixed $s$
and $r\to1$, the amplitude decays as $Pm_1\e^{-Pm_1}$, where
$P\propto(1-r)s/(1+s)$ as $m_1\to\infty$. This behavior is evident
for $r=0.3$ in  Fig.~\ref{fig2}(a), which shows that the divergence,
while obvious and dominant for $m_1=m_2<3$, rapidly becomes no more
than a minuscule spike, which soon becomes invisible on any
graphical plot. On the other hand, for greater values of the
coupling ratio $r$ the logarithmic divergence remains dominant
for larger values of $m_1$ and $m_2$ as seen in Fig.~\ref{fig2}(b).
But returning to Fig.~\ref{fig2}(a) with $r=0.3$, one observes
that as soon as the strip widths, $m_1=m_2$, exceed three lattice
spacings, there appear two further specific heat peaks, albeit
rounded; these grow rapidly in height and sharpness, and as $m_1$
and $m_2$ increase, they soon dominate the plots.

Now Fig.~\ref{fig2} is based on exact analytic calculations
expounded in Part II of this article.\cite{HAY} In fact, the
analysis of the finite-size behavior of planar Ising models
based on the exact solution of Onsager, as extended by
Kaufman,\cite{Kaufman} goes back to the work of Fisher and
Ferdinand\cite{FerdFisher,MEFJp} in 1969. Specifically, the
solubility of arbitrarily layered planar Ising models was first
noted and reported at a conference in Japan,\cite{MEFJp} while,
independently, McCoy and Wu\cite{MWbk} developed and analyzed
{\it randomly} layered Ising models. The thermodynamics for
regularly layered models was developed by Au-Yang and
McCoy\cite{HAYBM} and Hamm,\cite{Hamm} while the scaling
behavior of a single strip of finite width was elucidated by
Au-Yang and Fisher.\cite{HAYFisher}
 
In general the bulk critical temperature can be simply stated,
for a layered distribution as \cite {MEFJp}
\be
k_BT_c\langle\!\langle\ln \coth(J_x/k_BT_c)\rangle\!\rangle=
2\langle\!\langle J_y\rangle\!\rangle,
\label{tcg}\ee
where the brackets $\langle\!\langle\cdot\rangle\!\rangle$
denote an average over the distribution, random or regular of
the distinct number (say $n<\infty$) of lattice spacings
constituting a layer of finite width. For the alternating
layered Ising model, this becomes
\ba
&&2J_1(1+r s)\nonumber\\
&&=k_BT_c[\ln \coth (J_1/k_BT_c)+s\ln \coth (rJ_1/k_BT_c)],
\label{tc}\ea
which depends only on the weakness ratio $r$ and the relative
separation $s$.
 
Then as $m_1$ and $m_2$ become large, the upper and lower rounded
peaks approach limiting values, $T_{1c}$ and $T_{2c}$ (as evident
in Fig.~\ref{fig2}(a)), which, in fact, match the corresponding
bulk (i.e., uniform) two-dimensional Ising models with coupling
constants $J_1$ and $J_2$.  Thus the limiting values $T_{1c}$
and $T_{2c}(r)$ are known \cite{Kaufman, FerdFisher, MWbk} and
given by
\ba
&&k_BT_{1c}/J_1\simeq 2.269185312,\nonumber\\
&&k_BT_{2c}/J_1\simeq r\cdot 2.269185312.
\label{vtc12}\ea
It proves easy to establish the expected inequalities 
\be
T_{2c}(r)\le T_{c}(r,s)\le T_{1c}.
\ee
\section {Qualitative Observations}
\begin{figure*}[htbp]
\centering
\includegraphics[width=1\textwidth]{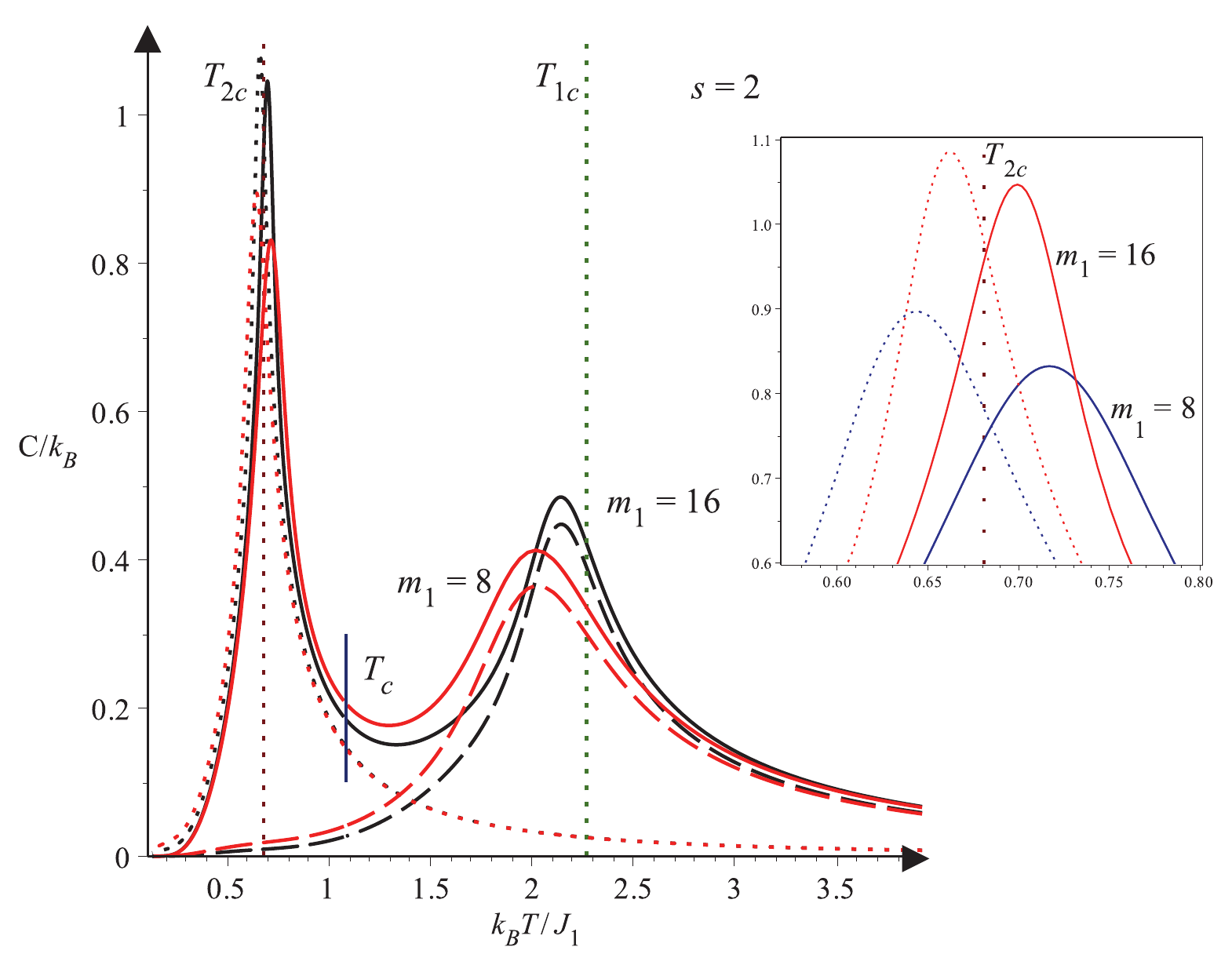}
\caption{(Color online) The specific heats of alternating layered
Ising models with relative strength $r=\!0.3$ (as in Fig.~\ref{fig2}
(a)) and relative separation $s=2$ for $m_1=8,16$. The dashed plots
denote the corresponding decoupled specific heats when $J_2=0$
($r=0$), while the dotted lines represent the specific heats when
$J_1=0$ (or $r\to\infty$). The inset displays the distinct maxima
near $T_{2c}$, dotted below but solid above.}
\label{fig3}
\end{figure*}

To explore further and develop the analogies with the observations
on superfluid helium systems, we retain the value of the weakness
ratio $r=0.3$ (used in Fig.~\ref{fig2}(a)) but increase the
relative layer separation to $s=2$. The results for $m_1=8$
and 16 (as used in Fig.~\ref{fig2}(a)) are presented in
Fig.~\ref{fig3}: see the solid curves. As anticipated, no sign
of any singularity at $T_c$ is visible. It should be noted,
nonetheless, that were one to examine the overall spontaneous
magnetization, $M_0(T)$, one would find --- and on a plot
see --- that $M_0$ vanished identically for $T>T_c$ but was
nonzero (and varying as $\propto (T_c-T)^\beta$ with
$\beta={\textstyle \frac 18}$ for 2D Ising
layers\cite{Fisher,Barber,MWbk}) as soon as $T<T_c$. In the
experiments on superfluids the analogous statement concerns
the overall superfluid density $\rho_s(T)$;\cite{GKMD} this
vanishes identically above the overall or bulk lambda transition
at $T_\lambda(\equiv T_c)$ but is detectable, via setting up
persistent superflow fluid currents, below
$T_\lambda$.\cite{PKMGn,PG} (In a bulk 3D superfluid
$\rho_s(T)$ varies as $(T_\lambda-T)^\zeta$ with
$\zeta\simeq0.67$, but in a planar 2D superfluid film of
thickness $L$, $\rho_s(T)$ increases discontinuously at the
corresponding superfluid transition temperature, $T_c(L)$,
on lowering the temperature.\cite{GKMD})

On the other hand, the temperatures of the upper and lower
rounded maxima increase (and decrease, respectively), as $m_1$
increases in Fig.~\ref{fig3}.  But now, using the explicit
results for the infinite strip of finite width,\cite {HAYFisher}
we also show, as dashed curves in Fig.~\ref{fig3}, the totally
decoupled $r=0$ (or $J_2=0$) plots for the two cases $m_1=8$
and $16$. Clearly the uncoupled upper maxima fall below $T_{1c}$
just as do the coupled ($r=0.3$) results. (It is worth remarking,
however, that for a finite $n\times n$ Ising lattice with periodic
boundary conditions, as studied by Ferdinand and
Fisher,\cite{FerdFisher} the maxima in the specific heats lie
above the bulk critical temperature $T_{1c}$.) Nevertheless,
there is clear evidence of a {\it coupling} or {\it proximity}
effect in that the specific heats for the alternating, coupled
system lie markedly {\it above} those for the decoupled ($r=0$)
strips. This same effect is seen in the experiments when finite
boxes are coupled by a helium film.\cite{PKMGn,PG}

\begin{figure*}[htbp]
\centering
\includegraphics[width=\textwidth]{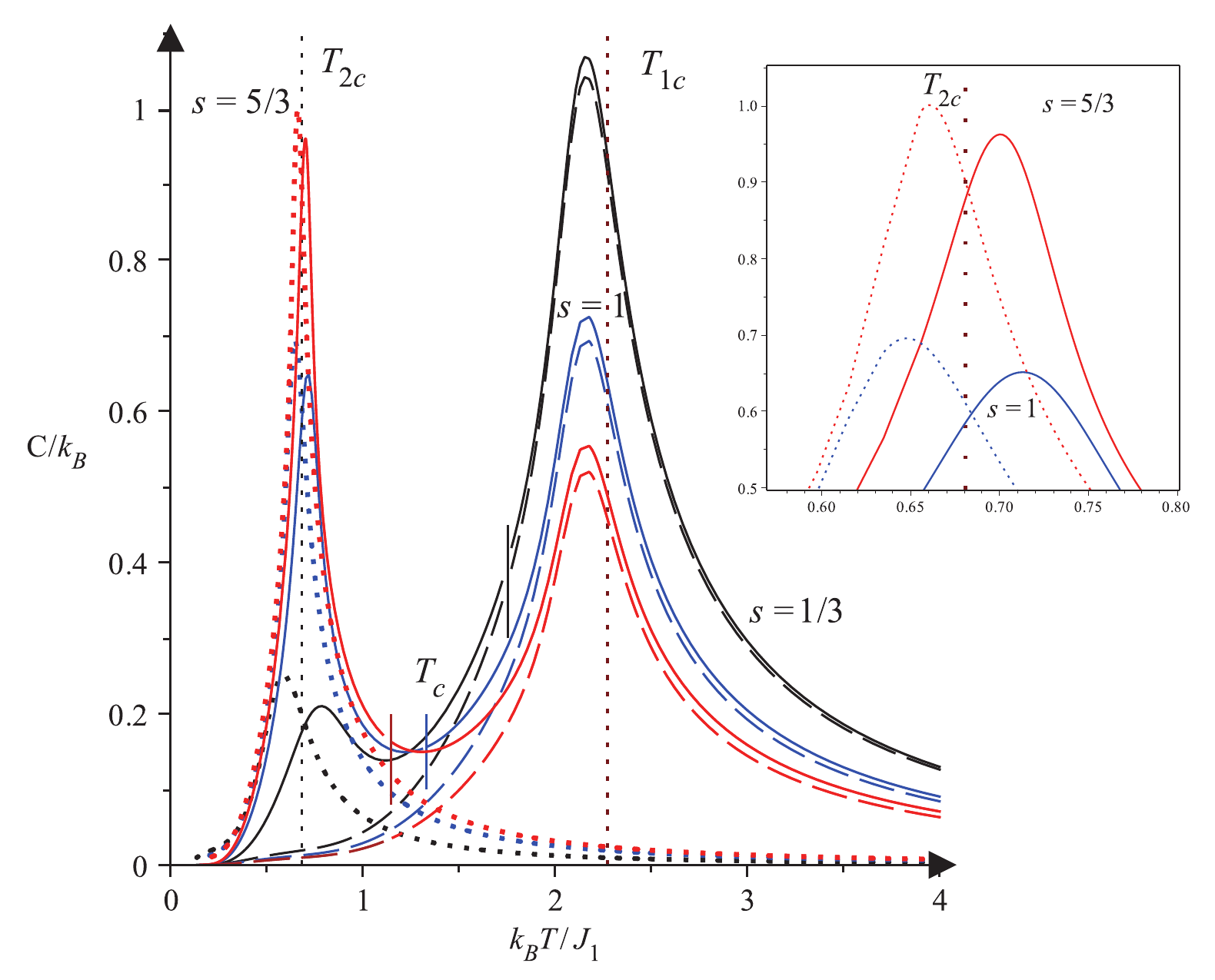}
\caption{(Color online) Specific heat plots for relative strength
$r=0.3$ (as in Figs.~\ref{fig2}(a) and \ref{fig3}) but with relative
separations $s=1/3,1$ and $5/3$ when $m_1=18$. Again the decoupled
$r=0$ behavior is seen in the dashed plots while for the opposite
limit, $r\to\infty$, the plots are dotted. The short vertical lines
locate the bulk critical points, $T_c(r,s)$, which decrease as $s$
increases. The inset shows various coupled and uncoupled maxima near
$T_{2c}$. }
\label{fig4}
\end{figure*}

Complementary phenomena are observed around the lower maxima. Thus
the dotted plots in Fig.~\ref{fig3} show the finite-width result
for the situation $r\to\infty$, or, more intuitively, $J_1=0$, for
$m_1=8$ and $16$ (i.e., $m_2=16$ and $32$). These decoupled
specific heats appear as very sharp, but still finitely rounded,
spikes. However, it must be noted that these $r\to\infty$ maxima
lie {\it below} $T_{2c}$, in accord with expectation for a
finite-width strip. On the other hand, the maxima of the coupled
alternating system lie {\it above} the limiting value $T_{2c}$ as
seen clearly in the inset in Fig.~\ref{fig3}. Once again there is
an unmistakable proximity or enhancement effect that is found also
in the experimental studies.\cite{PKMGn,PG}

As a next step of our qualitative exploration, we present in
Fig.~\ref{fig4} the effects of varying the relative separation
$s$ for significantly wide, $m_1=18$, strips spaced apart by
weaker strips of relative strength $r=0.3$ (as before). In this
case the first point to notice is that $T_c(r,s)$ increases quite
rapidly towards $T_{1c}$ as the separation $s$ approaches zero.
Next, the uncoupled ($r=0$) specific heats near $T_{1c}$ (shown
dashed as in Fig.~\ref{fig3}) all have maxima located at the same
temperature, determined only by $m_1=18$ for a finite width strips,
while their magnitude is determined by $s$ simply via normalization,
either through relative area or on a  per-site basis. However,
there is still clear enhancement in the coupled layers even though
the corresponding rounded maxima deviate very little in location
from the uncoupled case. By contrast near $T_{2c}$, as illustrated
by the inset, the displacements even of the uncoupled maxima
(shown dotted), depend significantly on the relative separation
ratio $s$.  Again, nonetheless, there is proximity induced
enhancement of the peaks both in magnitude and displacement
above $T_{2c}$.

Finally, we may enquire about the level of the specific heats
around the bulk critical point $T_c$ or in the vicinity of the
minima observed in Figs.~\ref{fig2}-\ref{fig4} that lie roughly
at $T_{min}\lesssim\halfs(T_{1c}+T_{2c})$. One may ask, for
example, how well the levels are approximated by appropriately
 weighted sums of the uncoupled peaks around $T_{1c}$ plus some,
perhaps reversed contribution from $T_{2c}$. For these purposes,
however, we need to proceed more quantitatively.

\section{ Scaling Explorations Near The Maxima}
We would like to relate the observations embodied in
Figs.~\ref{fig2}-\ref{fig4} to more general scaling concepts.
To this end, recall\cite{Fisher,Barber,MEFJp,FerdFisher}
that a bulk system with a critical temperature $T_c$ may be
characterized by a correlation length $\xi(T)$ which diverges
on approach to criticality as
\be
\xi(T)\approx\xi_0/|t|^\nu\quad \hbox{with}\quad t=(T/T_c)-1\to0,
\ee
where $\nu$ is a characteristic critical exponent while $\xi_0$
is a length of order the lattice spacing $a$, or molecular size,
etc. For 2D Ising systems one has\cite{Fisher,Barber,FerdFisher,MWbk}
$\nu=1$, whereas for superfluid helium in three bulk dimensions
$\nu\simeq 0.67$.\cite{GKMD} Then in a system limited in size by a
finite length $L=\ell a$, the scaling hypothesis asserts, in general
terms, that when $\ell$ and $\xi(T)$ are large enough, the rounding
of critical point singularities is primarily controlled by the
ratio $y=L/\xi(T)$.

Consequently, for the finite-size behavior of the specific heat
per site, which diverges in bulk as $|t|^{-\alpha}$ where $\alpha$
is typically small (or even negative), the basic scaling hypothesis
may be expressed as 
\be C(\ell; T)\approx \ell^{\alpha/\nu}[Q(x)-Q_0]/\alpha,
\label{scalingC}\ee
where $Q(x)$ is the scaling function while the scaled temperature is 
\be x=\ell^{1/\nu}t\propto y^{1/\nu}=[L/\xi(T)]^{1/\nu},
\ee
and $Q_0>0$ is a constant parameter. The exponent $\alpha$ in the
denominator in (\ref{scalingC}) allows for the limit $\alpha\to0$,
which yields, with $Q(0)\to Q_0$, a logarithmic singularity as is
appropriate for 2D Ising systems. One may then take
\be
C(\ell;T)\approx(Q_0/\nu)\ln\ell+Q(\ell^{1/\nu}t)
\label{IsingSc}\ee
as the basic hypothesis where, for use below, we note that at
criticality one has $C(\ell;T_c)\approx(Q_0/\nu)\ln\ell+Q(0)$.
In fact, this hypothesis has been established explicitly for
infinite Ising strips of width $\ell$ and $Q(x)$ has been
explicitly determined.\cite{HAYFisher}

\subsection{Upper Maxima near ${\bf T_{1c}}$}
\begin{figure}[htbp] 
\centering
\includegraphics[width=\hsize]{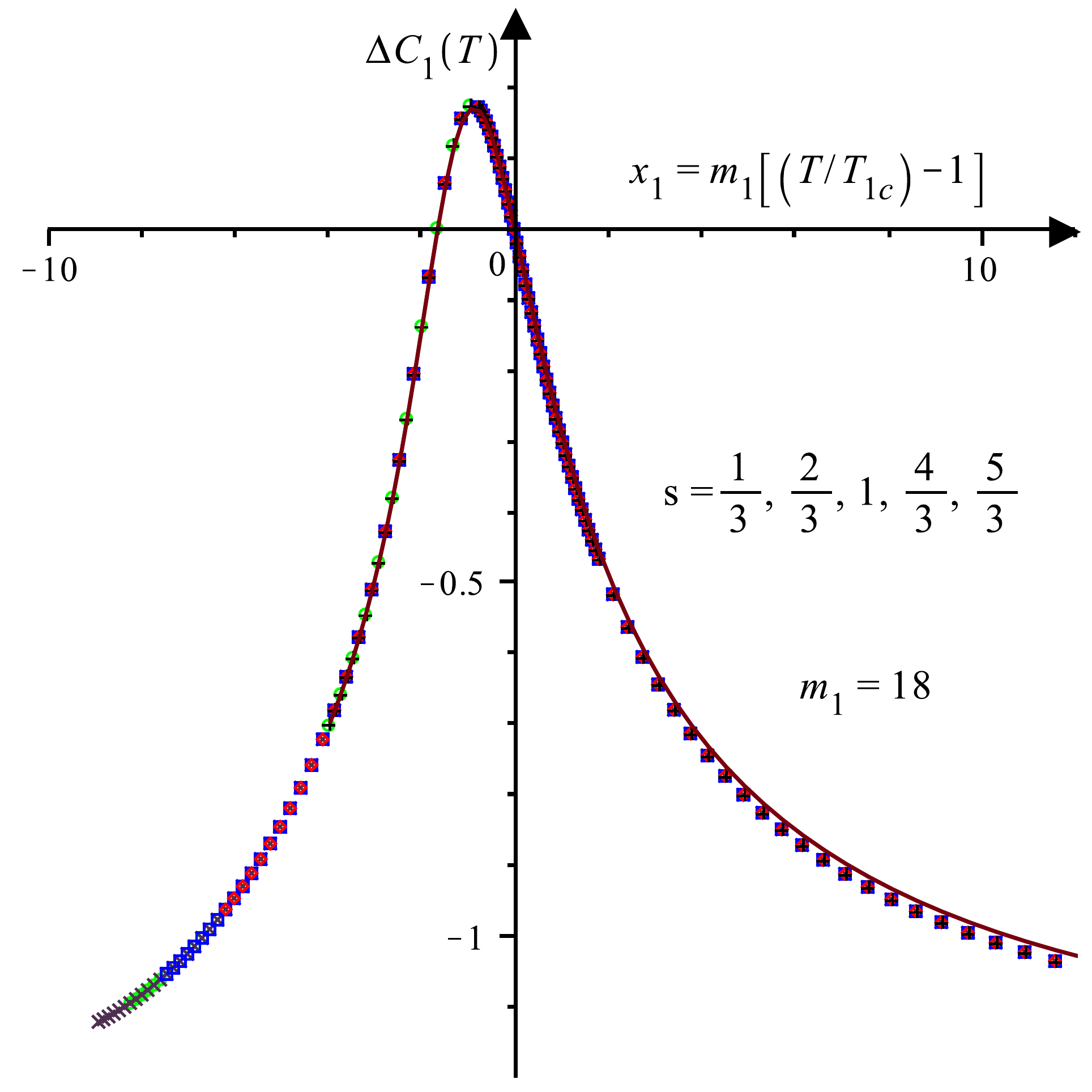}
\caption{(Color online) Scaling plot vs $x_1=m_1t_1$ for the upper
or stronger net finite-size contribution, namely,
$\Delta{C}_1(J_1,J_2;T)$ as defined in the text for strong strips
of width $m_1=18$ at various separations but fixed $r=0.3$.
The solid curve is the plot of the specific heat of an infinite
strip of width $m_1=18$ and coupling $J_1$ when its value at the
bulk critical temperature $T_{1c}$ is subtracted.\cite{HAYFisher}}
\label{fig5}
\end{figure}

To apply these concepts to our layered Ising system, in the first
case for the upper maxima near $T_{1c}$, we recall from
Figs.~\ref{fig3} and \ref{fig4} that leaving aside relatively
small enhancements in magnitude, the total specific heat,
$C(J_1,J_2;T)$, approaches rather well the limiting
forms\cite{HAYFisher} of a suitably normalized single strip of
width $m_1$. Accordingly, we subtract a contribution from
non-coupled weaker Ising strips by defining
\be
{C_1}(J_1,J_2;T)=(1+s)[C(J_1,J_2;T)-C(0,J_2;T)],
\label{C1}\ee
where the normalization factor $(1+s)$ is needed for the scaling
plots now to be examined. Finally in accord with (\ref{IsingSc})
and the subsequent remark we introduce the upper or stronger
{\it net finite-size contribution}
\be
\Delta{C_1}(J_1,J_2;T)=C_1(J_1,J_2;T)-C_1(J_1,J_2;T_{1c}),
\label{dC1}\ee
in which the value at the limiting critical point, $T_{1c}$,
has been subtracted. If we accept the identifications
$\ell\Rightarrow m_1$ and $t\Rightarrow t_1=(T/T_{1c})-1$ and
recall $\nu=1$, we might expect $\Delta C_1(T)$ to obey scaling
in terms of the scaled temperature variable
\be
x_1=m_1[(T/T_{1c})-1]\equiv m_1t_1.\label{x1}
\ee
This expectation is well supported by the plots in Fig.~\ref{fig5}
for $m_1=18$ and $s=n/3$ for $n=1,2,\cdots,5$:
the ``data collapse'' is strikingly well realized. 

Beyond this, however, explicit calculations \cite{HAY} show that,
asymptotically, $\Delta C_1(T)$ is simply related to the limiting
scaling function, $Q^\infty(x_1)$, for an infinite strip of
coupling $J_1$ and width $m_1$ already known
explicitly.\cite{HAYFisher} Specifically, allowing for
normalization, yields
\ba
&&\Delta{C_1}(J_1,J_2;T)\approx(1+s)[C(J_1,0;T)-C(J_1,0;T_{1c})]
\nonumber\\
&&\qquad\approx Q^\infty(x_1)-Q^\infty(0),
\label{dC1sc}\ea  
where, to complete the description we report\cite{HAY}
\ba
&&(1+s)C_1(J_1,J_2;m_1,m_2;T_{1c})\approx C^\infty(J_1;m_1;T_{1c})
\nonumber\\
&&\qquad\approx A_0\ln m_1+Q^\infty(0); \nonumber\\
&&A_0=2[\ln(\sqrt2+1)]^2/{\pi},\quad Q^\infty(0)\approx0.30681A_0,
\label{Q0}\ea
which (recognizing that $\nu=1$ for 2D Ising models) is in accord
with (\ref{IsingSc}). Note that in this limit not only has the
dependence on $m_2$ dropped out but also the dependence on $J_2$.
However, as regards the enhancement seen in
Figs.~\ref{fig2}-\ref{fig4}, we know that $m_2$ and $J_2$
do play a role. This will be studied further below.

\subsection{Lower Maxima near ${\bf T_{2c}}$}
Let us now shift attention to the behavior of the specific heat
peaks of the alternating system, near the lower (or weaker)
limiting critical point, $T_{2c}$. The rounded maxima are shown
in detail in the insets of Figs.~\ref{fig3} and \ref{fig4}. Now we
can follow the procedure that led to the definition (\ref{C1}).
Thus we consider the normalized difference
\be
{C_2}(J_1,J_2;T)=(1+s^{-1})[C(J_1,J_2;T)-C(J_1,0;T)].
\label{C2}\ee
Then, following again the previous analysis, the {\it weaker net
finite-size contribution} may be defined as in (\ref{dC1}), by
\be
\Delta{C_2}(J_1,J_2;T)=C_2(J_1,J_2;T)-C_2(J_1,J_2;T_{2c}).
\label{dC2}\ee
It is natural to suppose that $\Delta C_2(T)$ might obey scaling
in terms of the new scaled temperature variable
\be
x_2=m_2[(T/T_{2c})-1]\equiv m_2t_2.
\ee
\begin{figure}[htbp]
\centering
\includegraphics[width=\hsize]{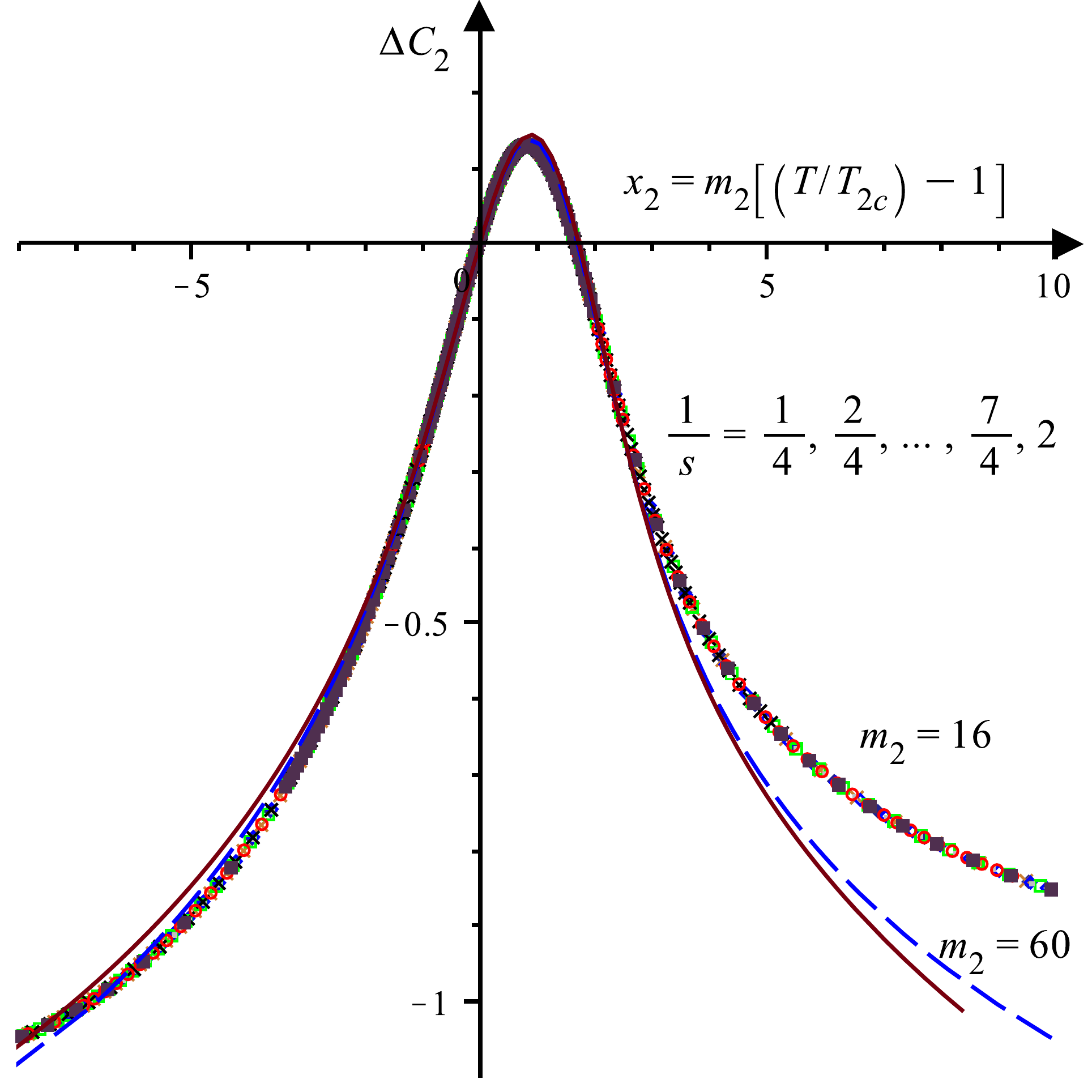}
\caption{(Color online) Scaling plots versus the scaling variable
$x_2=m_2t_2$ for the lower or weaker net finite-size contribution
$\Delta{ C}_2(J_1,J_2;T)$, as defined in relations
(\ref{C2})-(\ref{dC2}), for strips of width $m_2=16$ and $r=0.3$
for various relative separations. The solid curve represents the
corresponding asymptotic form $Q^\infty(-x_2)-Q^\infty(0)$, for
an infinite strip of finite width and coupling $J_2$, with the
temperature {\it reflected} about $T=T_{2c}$.\cite{HAYFisher}
The dashed curve represents data for a wider strip with $m_2=60$.}
\label{fig6}
\end{figure}
This hypothesis is tested in Fig.~\ref{fig6} and, evidently, is
remarkably successful, exhibiting excellent data collapse. But
more remarkable yet is the evidence provided by the solid line
plotted in Fig.~\ref{fig6}. This derives directly from the limiting
scaling function for an infinite strip\cite{HAYFisher} of
coupling $J$ and finite width $m$ but with the sign of the argument
{\it reversed}. In other words, the previous asymptotic form
(\ref{dC1sc}) is now, as established in Part II,\cite{HAY} replaced by
\ba
&&\Delta{C_2}(J_1,J_2;T)\approx
(1+s^{-1})[C_2(0,J_2;{\checkT})-C_2(0,J_2;T_{2c})]
\nonumber\\
&&\qquad\approx Q^\infty(-x_2)-Q^\infty(0),
\label{dC2sc}\ea
where the modified temperature ${\checkT}$ is simply attained by
reflecting about $T_{2c}$; explicitly we have
\be {\checkT}(T)= T_{2c}-(T-T_{2c})=2T_{2c}-T.
\label{checkT}\ee
It is appropriate to recall (\ref{Q0}) which may now be rewritten
to complement (\ref{dC2sc}) as
\ba
&&(1+s^{-1})C_2(J_1,J_2;m_1,m_2;T_{2c})\approx
C^\infty(J_2;m_2;T_{2c})\nonumber\\
&&\qquad\approx A_0\ln m_2+Q^\infty(0),
\label{Q2}\ea
in which the values of $A_0$ and $Q^\infty(0)$ are given in (\ref{Q0}).
We may note, further, that in this limit the original dependence on
both $J_1$ and $m_1$ has vanished; but, once more, there are clear
residual effects associated with the proximity and interlayer couplings.

\section{Enhancement Effects}
To address the behavior of the specific heats beyond the leading
scaling behavior revealed in Figs.~\ref{fig5} and \ref{fig6}, we
may define an ``enhancement'' by subtracting from the total specific
heat per site contributions deriving from the corresponding
independent uncoupled strips. However, in doing this we must
recognize --- following Fig.~\ref{fig6} and the result
(\ref{dC2sc}) --- that a reversed or reflected temperature variable
is needed around $T_{2c}$. To this end we utilize the modified
temperature variable, ${\checkT}(T)$, defined in (\ref{checkT}).
Thus we specify the net {\it enhancement} for fixed $m_1$ and $m_2$ by
\ba
&&{\cal E}(J_1,J_2;m_1,m_2; T)\nonumber\\
&&\quad=C(J_1,J_2; T)-C(J_1,0; T)-C{\mbox{\boldmath$($}}0,J_2;
{\checkT}(r){\mbox{\boldmath$)$}}.
\label{enhancement}\ea
\begin{figure}[htbp]
\centering
\includegraphics[width=\hsize]{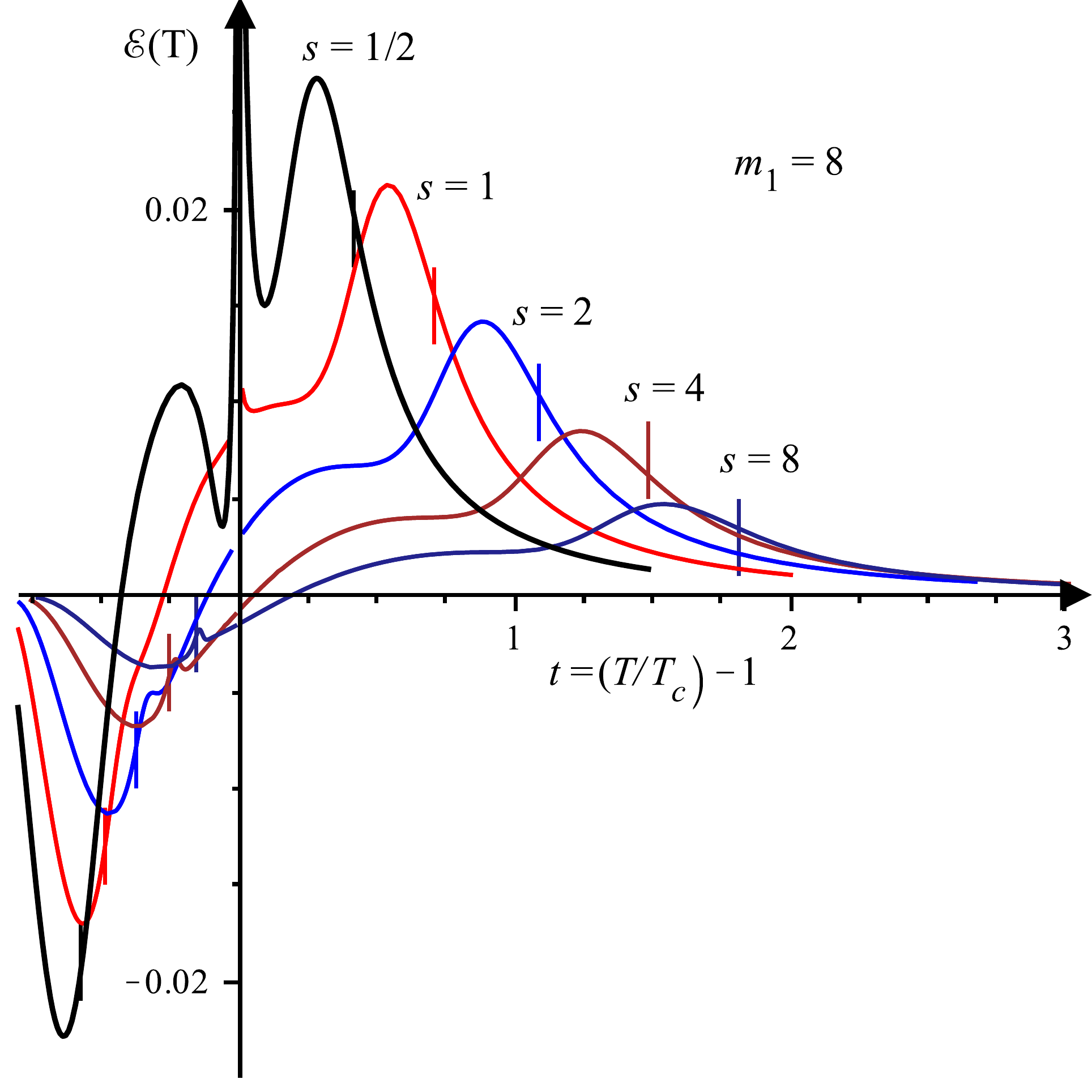}
\caption{(Color online) Plots of the enhancement ${\cal E}(t)$
versus $t=(T/T_c)-1$ for $m_1=8$, $r=0.3$ and various relative
separations $s$. The short vertical lines above $T_c(r,s)$, i.e.,
for $t>0$, are the corresponding positions of $T_{1c}$, while below
$T_c(r,s)$, they locate $T_{2c}$.}
\label{fig7}
\end{figure}

It is worth remarking parenthetically that in adopting this
definition of the enhancement we are, in particular, utilizing the
theoretical result (\ref{dC2sc}) proved for the alternating Ising
strips.\cite{HAY} In more general situations (such as confined
superfluid helium) the last term in (\ref{enhancement}) should be
replaced by an asymptotic term obtained through an appropriate
initial data analysis of the behavior close to $T_{2c}$ such as
led to the original (finite $m_2$) form in Fig.~\ref{fig6}.

In Fig.~\ref{fig7}, we plot the enhancement for our alternating Ising
strips with $r=0.3$ as a function of $t=[(T/T_c)-1]$ for $m_1=8$ and
various separations $s$. One sees that the logarithmic divergence
at $t=0$ is barely visible for $s=1$, and essentially disappears for
$s>1$. In addition, as expected, the magnitude of the enhancement
decreases as $s$ (or $m_2$) increases; but by what law?

\begin{figure}[htbp]
\centering
\includegraphics[width=\hsize]{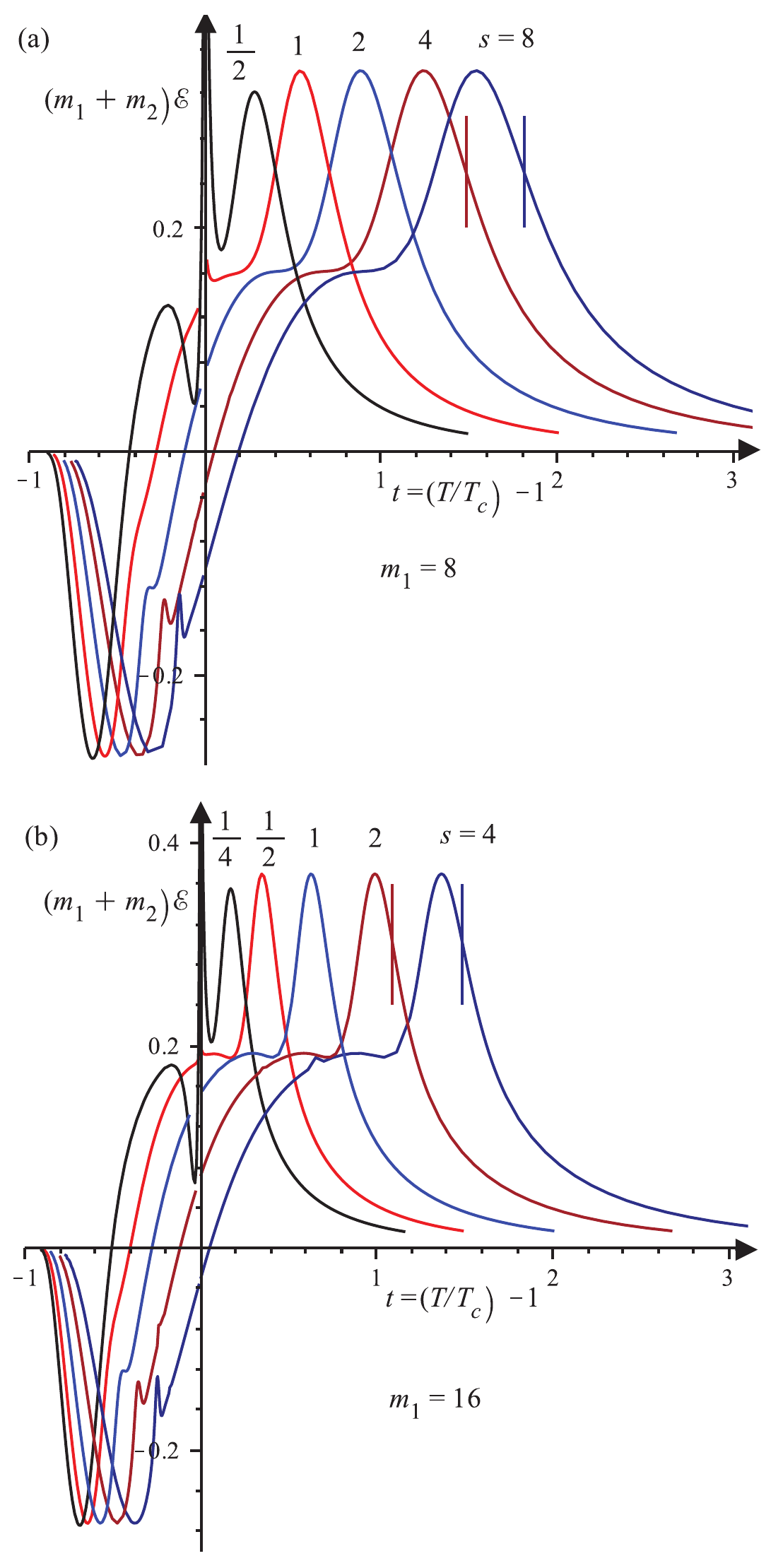}
\caption{(Color online) The enhancement ${\cal E}(t)$ multiplied by
$(m_1+m_2)$: (a) for $m_1=8$ as in Fig.~\ref{fig7}, and
(b) for $m_1=16$. The short vertical lines locate the corresponding
upper limiting critical points, $T_{1c}$.}
\label{fig8}
\end{figure}

To address this question we recall, first, that the leading
correction to the asymptotic form of the specific heat of an
infinite strip of finite width $m$ must arise from the two
non-vanishing {\it boundary free energy contributions}%
\cite{FerdFisher,MWbk,HAYFisher,FisherFerd} which yield a total
specific heat term of relative order $1/m$. The effects of this are
already evident in Fig.~\ref{fig6} where the primary contribution
(solid curve) is, especially for $x_2\geq 2$, more closely approached
by the data for $m_2=60$ than that for $m_2=16$. It is clear that
such corrections must arise also in the bulk alternating strip system
from the regularly spaced modified boundaries or {\it seams}. By
the same token, boundaries or surface effects play similar roles
in the experiments on the dimensional crossover behavior of bulk
specific heats of helium\cite{GKMD,KMG} and should enter to some
degree also for small helium boxes  coupled via helium films,
etc.\cite{PKMGn,PG,PKMGpr}

Accordingly, Fig.~\ref{fig8} presents the enhancements ${\cal E}(t)$
versus $t\propto[T-T_c(r,s)]$, but now multiplied by the factor
$(m_1+m_2)$ which clearly should account in leading order for the
density of seams in the bulk. It is striking that the maxima
(close to $T_{1c}$) and the minima (near $T_{2c}$) appear to
rapidly approach almost constant values. This represents strong
evidence that the enhancement ${\cal E}(J_1,J_2;m_1,m_2;T)$ is
of order $1/(m_1+m_2)$ as the relative separation, $s=m_2/m_1$,
increases at fixed $m_1$.

However, by comparing Figs.~\ref{fig8}(a) and \ref{fig8}(b), it
becomes clear that the behavior of the rescaled enhancement peaks
that approach $T_{1c}$, when $s$ increases, depend quite noticeably
on $m_1$, the width of the strong strips. Specifically, the
enhancement peaks become both narrower, as indeed implied by
Fig.~\ref{fig6}, and taller as $m_1$ grows. 

Consequently, we will separately investigate the behavior of the
enhancement close to $T_{1c}$, noting that some logarithmic
dependence on $m_1$ might be present; in complementary fashion
there might be a logarithmic variation with $m_2$ in the vicinity
of $T_{2c}$. Nevertheless, Fig.~\ref{fig8} suggests that the
enhancements rescaled by $(m_1+m_2)$ might approach more or less
constant shapes in the interval $T_{2c}<T<T_{1c}$.
 
 Then, since the expected scaling behavior must switch in the
region between $T_{1c}$ and $T_{2c}$, we anticipate, on the one
hand, that the rescaled enhancement near $T_{1c}$ as functions
of $t_1=(T/T_{1c})-1$ are independent of $m_2$ in accord with
the data collapse seen in Fig.~\ref{fig5}, while on the other
hand, near $T_{2c}$ the rescaled enhancements as functions of
$t_2=(T/T_{2c})-1$ depend on $m_2$ but become independent of
$m_1$, as borne out by Fig.~\ref{fig6}.
 
\begin{figure*}[htbp]
\centering
\includegraphics[width=0.93\textwidth]{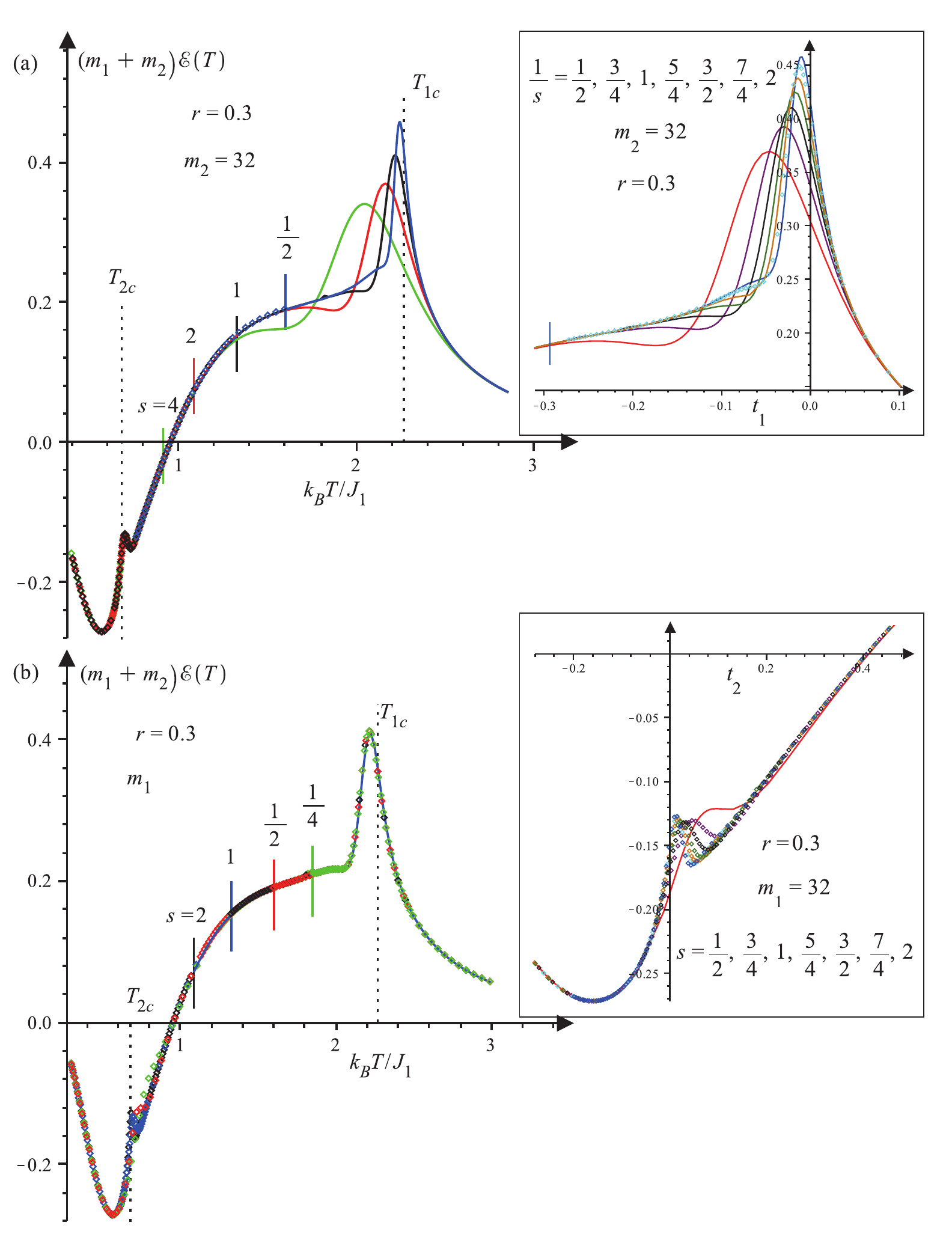}
\caption{(Color online) The rescaled enhancement
$(m_1+m_2){\cal E}(T)$ for $r=0.3$ is plotted for $m_2=32$,
and $m_1=8,16,32,64$ in (a) showing that data collapses occur
near $T_{2c}$. The framed inset shows more detail near $T_{1c}$
as a function of $t_1=(T/T_{1c})-1$. In (b) the plots are
for $m_1=32$, and $m_2=8,16,32,64$. Now the data become independent
of $m_2$ near $T_{1c}$. The behavior near $T_{2c}$ is plotted versus
$t_2=(T/T_{2c})-1$ in the frame.}
\label{fig9}
\end{figure*}

\begin{figure*}[htbp]
\centering
\includegraphics[width=0.93\textwidth]{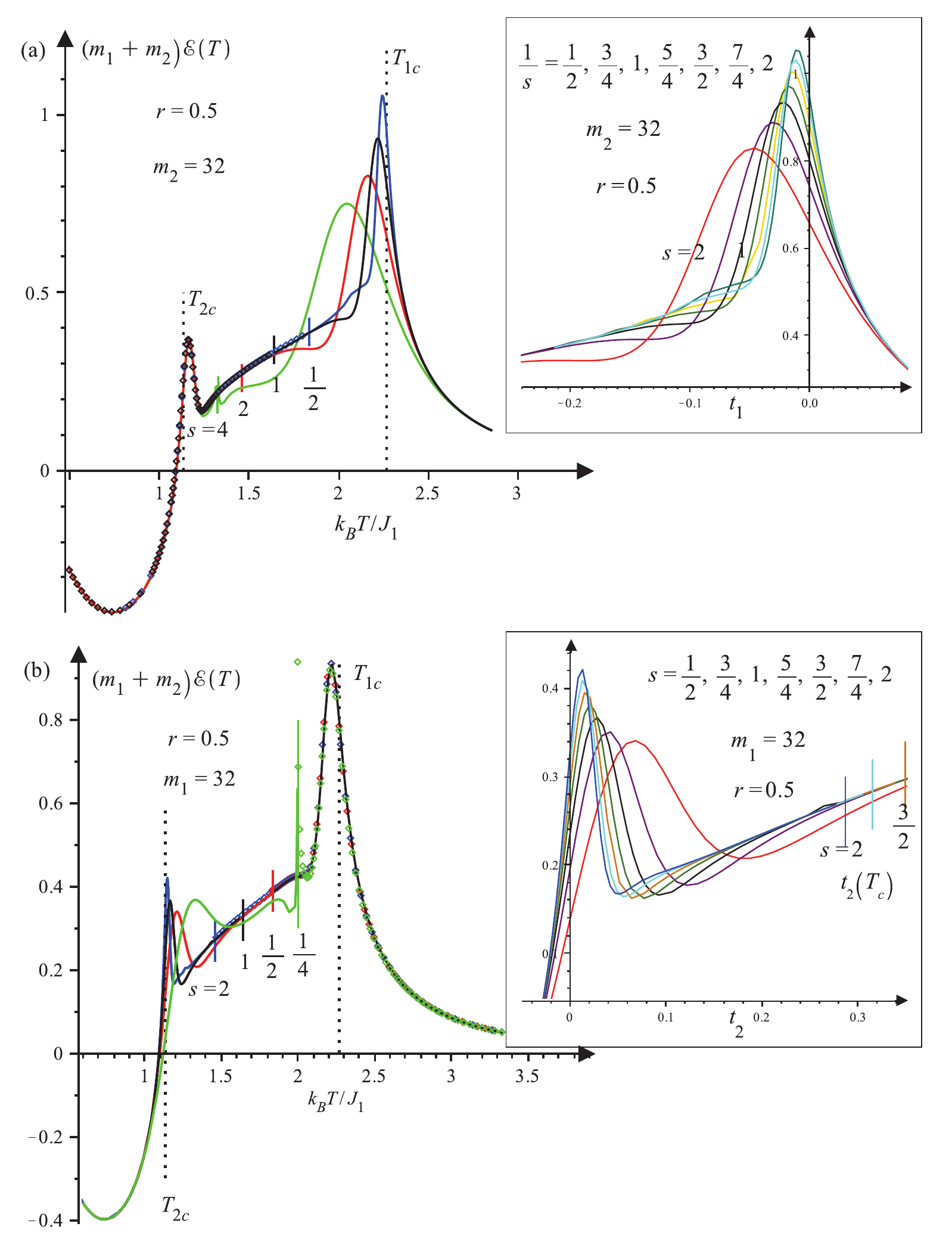}
\caption{(Color online) Plots of the rescaled enhancement
$(m_1+m_2){\cal E}(T)$ for $r=0.5$ as in Fig.~\ref{fig9}.
The short vertical lines denoted the positions of $T_{c}(s)$.}
\label{fig10}
\end{figure*}

\begin{figure*}[htbp]
\centering
\includegraphics[width=0.90\textwidth]{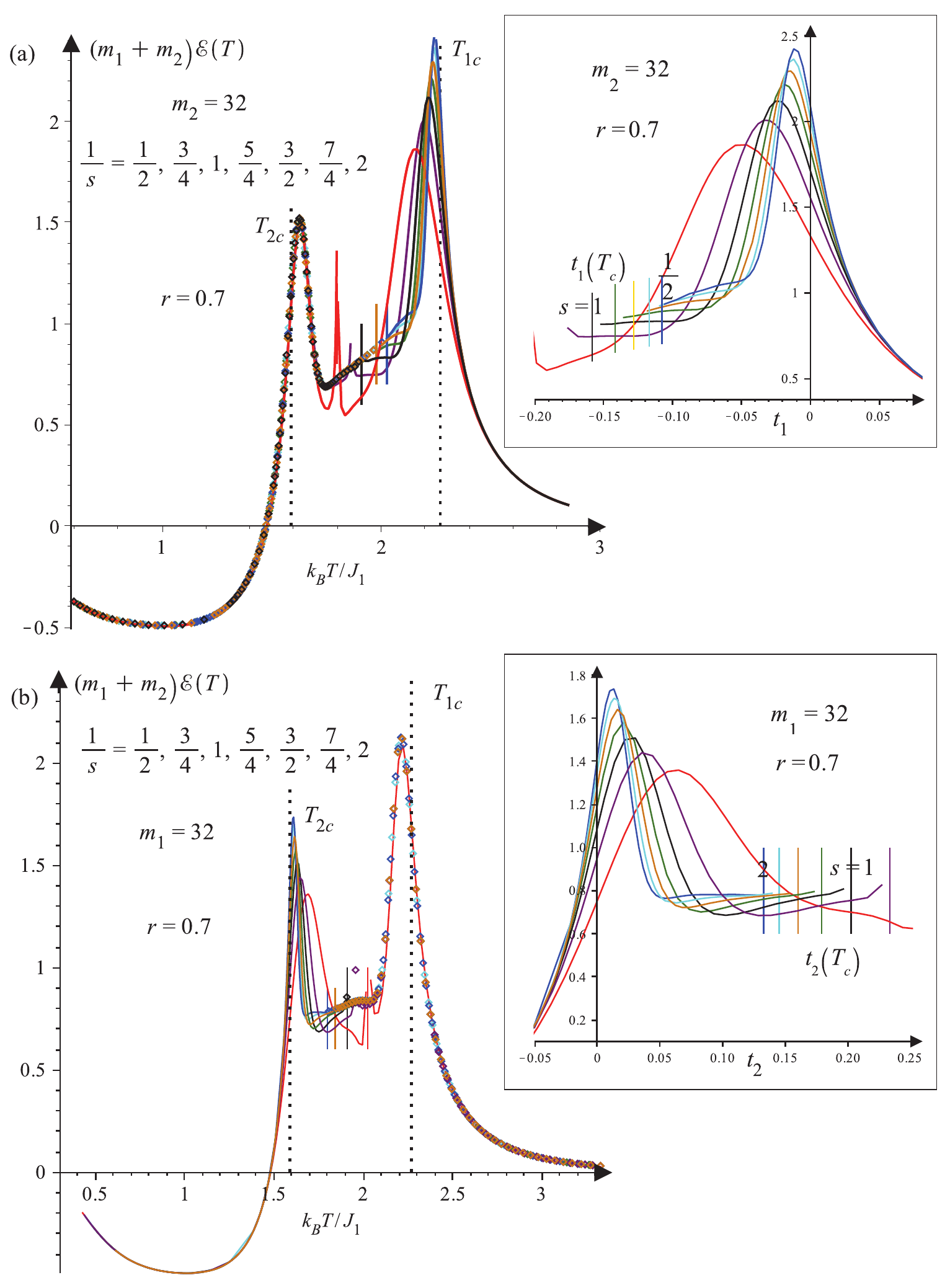}
\caption{(Color online) The rescaled enhancements
$(m_1+m_2){\cal E}(T)$ for $r=0.7$: (a) for $m_2=32$, and
$m_1=8n$ for $n=2,3,\cdots,7, 8$; data collapse occurs near $T_{2c}$,
while the frame shows details near $T_{1c}$ vs. $t_1$;
(b) for $m_1=32$, and $m_2=8n$ for $n=2,3,\cdots,7, 8$;
the plots near $T_{1c}$ are now independent of $m_2$, while the
behavior near $T_{2c}$ is shown in the frame.}
\label{fig11}
\end{figure*}

Accordingly, in Figs.~\ref{fig9}-\ref{fig11} we plot the enhancements
rescaled by $(m_1+m_2)$ for the relative strengths $r=0.3$, $0.5$, and
$0.7$, respectively. 
In parts (a) of these figures, the plots are for fixed $m_2=32$, with
stronger strips of widths $m_1=8n$ for increasing values of $n(\le 8)$.
Evidently, the rescaled enhancements are close to independent of
$m_1$ for $T$ near $T_{2c}$. The framed plots in the figures present
more detail as a function of $t_1$.

In part (b) of Figs.~\ref{fig9}-\ref{fig11}, the widths of the
stronger strips are fixed at $m_1=32$, while $m_2=8n$ increases.
Now data collapse is seen near $T_{1c}$. In the frames the reduced
enhancement are plotted near $T_{2c}$ as functions of $t_2$ for the
increasing values of $m_2$.

Inspection of Figs.~\ref{fig9}-\ref{fig11} demonstrates that as
$m_1$ increases, the upper maxima approach $T_{1c}$ from below, and
grow steadily in height resembling the corresponding specific heats
shown in Fig.~\ref{fig2}(a). By contrast, the lower rounded peaks
of the rescaled enhancements, though much smaller, lie {\it above}
the limit $T_{2c}$ and similarly grow in height as $m_2$ increases.
These observations in comparison with Figs.~\ref{fig2}(a), \ref{fig3},
and \ref{fig4} and the subsequent scaling analyses utilizing relations
(\ref{IsingSc}), (\ref{Q0}) and  (\ref{Q2}), strongly suggest the
presence of a logarithmic dependence of the peak heights on $m_1$
for $T>T_c$, but on $m_2$ for $T<T_c$.

\begin{figure}[htbp]
\centering
\includegraphics[width=\hsize]{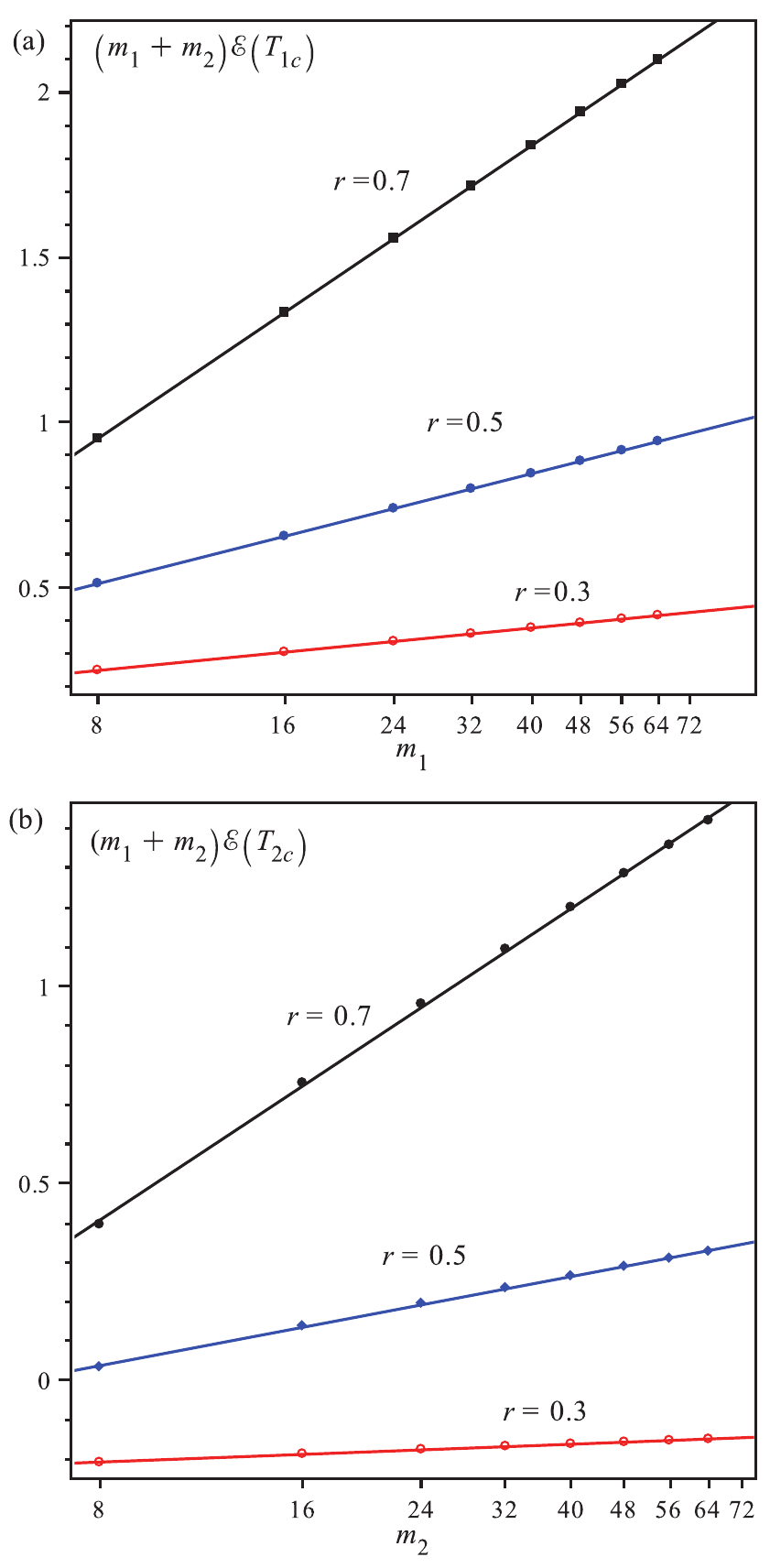}
\caption{(Color online) The rescaled enhancement
$(m_1+m_2){\cal E}(r;T)$ evaluated at the limits
(a) $T_{1c}$ and (b) $T_{2c}$ for three values of $r$,
plotted versus $\ln m_1$ and $\ln m_2$, respectively.
The linear fits are as specified in (\ref{logfit}) and Table I.}
\label{fig12}
\end{figure}

To investigate this issue concerning the vicinities of $T_{1c}$, and
$T_{2c}$ further, we have calculated the critical values of the
rescaled enhancements, namely, $(m_1+m_2){\cal E}(T_{1c})$ and
$(m_1+m_2){\cal E}_{2c}(T_{2c})$, for $r=0.3$, $0.5$, and $0.7$ and
for eight specific values of $m_1$ or $m_2$, respectively, in the range
8 up to 64. The results are plotted vs.\ $\ln m_{1,2}$ in
Fig.~\ref{fig12}.

Evidently the data are well described by the form 
\ba
(m_1+m_2){\cal E}^{\pm}_{c}(m)\simeq
{\cal B}^{\pm}(r)\ln[m/m^{\pm}_0(r)],\quad T_{ic}\gtrless T_c,
\nonumber\\
\label{logfit}\ea
where fitted values of the amplitudes, ${\cal B}^{\pm}(r)$,
and offsets, $m^{\pm}_0(r)$, for the upper and lower maxima,
are set out in Table I. Both the amplitude and
the offset appear to vary exponentially rapidly with $r$ in the
region, say $0.2<r<0.8$.

\begin{table}[htdp]
\begin{tabular}{|c|c|c|c|}\hline
&\quad $r=0.3$ \quad&\quad $r=0.5$ \quad&\quad $r=0.7$ \quad\\\hline
$\quad {\cal B}^{+}(r)\quad$&\quad0.0800\quad&\quad0.2071\quad&
\quad0.5115\quad\\\hline
$\quad m^{+}(r)\quad$&\quad0.358\quad&\quad0.677\quad&\quad1.427
\quad\\\hline
$\quad {\cal B}^{-}(r)\quad$&\quad0.0282\quad&\quad0.1405\quad&
\quad0.4913\quad\\\hline
$\quad m^{-}(r)\quad$&\quad$12600$&\quad6.13\quad&\quad3.50
\quad\\\hline
\end{tabular}
\caption{ Amplitudes and offsets for the rescaled enhancements at
$T_{1c}$ and $T_{2c}$ as shown in Fig.~\ref{fig12}. Note that the
very large value $m^{-}(0.3)=12600$ combined with the small value
for ${\cal B}^{-}(0.3)$ yields $ {\cal B}^{-}\ln(8/m^{-}_0)=-0.207$
which agrees with the plot in Fig.~\ref{fig12}(b).}
\label{default}
\end{table}

Beyond the relatively slow logarithmic growth of the enhancement
maxima at both limits, $T_{1c}$ and $T_{2c}$, it is reasonable, on
the basis of Figs.~\ref{fig9}-\ref{fig11}, to speculate as to the
limiting behavior of ${\cal E}(J_1,J_2; m_1,m_2; T)$ in the three
regions: above, below, and in-between $T_{1c}$ and $T_{2c}$.

It seems natural to propose, first, a logarithmic form in $t_1$
and $t_2$, say,
\ba
&(m_1+m_2){\cal E}(T)\hspace*{6em}&\nonumber\\
&\quad\approx\,\mathcal{A}^{+}(r)\ln|t_1|+\mathcal{C}^{+}(r),&
\text{ if $t_1>0$,}\nonumber\\
&\quad\approx\,\mathcal{A}^{-}(r)\ln|t_2|+\mathcal{C}^{-}(r),&
\text{ if $t_2<0$,}
\ea
as valid above and below $T_{1c}$ and $T_{2c}$. Since the limit
$r\to1$ corresponds to a uniform Ising square lattice with a
symmetric logarithmic singularity, as in (\ref{log}), it might be
tempting to guess that the amplitudes ${\cal A}^+(1)$ and
${\cal A}^-(1)$, and the backgrounds, ${\cal C}^+(1)$ and
${\cal C}^-(1)$, are equal; but that would surely go beyond what
our numerical evidence might support.

As regards the intermediate regions, however, a very different
behavior seems implied. Thus, ignoring the logarithmic spikes, for
$T$ between $T_{1c}$ and $T_{2c}$ and for $r$ not too large, the
enhancement $(m_1+m_2){\cal E}(r;T)$, appears to increase smoothly
and monotonically. Indeed the large $s^{-1}$ plots are almost linear.
On extrapolating this linearity up to $T_{1c}$ and down to $T_{2c}$
in a nonsingular fashion, one finds clear numerical limits for
$t_1\to0^-$ and $t_2\to0^+$. Specifically, the numerical evidence
suggests the increasing values
\ba
(m_1+m_2){\cal E}^+_c(r)\simeq &0.27,\quad& 0.56,\quad 1.3,\nonumber\\
(m_1+m_2){\cal E}^-_c(r)\simeq &-0.20,\quad& 0.15,\quad 0.75,
\ea
for $r=0.3$, $0.5$, and $0.7$, respectively. While further numerical
studies might reduce the uncertainties of these approximate estimates,
a firm theoretical base unfortunately seems beyond current reach. 

\section{SUMMARY : 2D-1D ISING vs. 3D-0D SUPERFLUID HELIUM}
    
In this Section we will summarize our study of connectivity and
proximity in two-dimensional alternating layered Ising models and
examine the relationships to the extensive studies of Gasparini
and coworkers\cite{GKMD,KMG,PKMG,PKMGn,PG,PKMGpr} on coupling and
proximity effects in small ``boxes'' of liquid helium-4 in the
vicinity of the bulk, three-dimensional superfluid transition.  

To start, we considered a set of strong square-lattice Ising model
strips, with spin-spin interaction $J_1$ and finite width $ m_1$,
that in the limit $m_1 \to\infty$ have a bulk two-dimensional Ising
transition with a logarithmically divergent specific heat at a
temperature $T_{1c}$. For finite $m_1$, however, an isolated
one-dimensional strip will display only a rounded maximum at a
lower temperature, say  $T_{1max}$, which, for $m_1$ large enough, 
will be well described by finite-size scaling
theory.\cite{Fisher,Barber} Our numerical studies explored values
of $m_1$ up to 64.

This situation may be compared with three-dimensional but finite-sized,
and hence zero-dimensional, ``boxes'' of liquid helium of linear
dimension, say  $L_1$, which in the limit  $L_1\to\infty$  will
exhibit a sharp specific heat singularity at the bulk lambda point,
$T_\lambda$. In the experiments of Gasparini and
coworkers,\cite{PKMG,PKMGn,PG,PKMGpr} box sizes $L_1=1\,\mu$m
and $2\,\mu$m were examined, as described further below. But it
might be noted that, on accepting a microscopic scale\cite{PKMGpr}
$\xi^+_0 = 0.143$ nm, these magnitudes of $L_1$ might more
realistically  be viewed as corresponding to $m_1\simeq7,000\,$-$14,000$,
values far beyond our computing capabilities.

Second, in the Ising context (as illustrated in Fig.~\ref{fig1})
the infinite number of strong strips were {\it connected } by weak
or coupling strips with interactions  $J_2=rJ_1$ (with $r <1$)
and width $m_2=sm_1$ [as introduced in (\ref{dfrs})], where our
exact calculations yielded explicit results for interaction ratios
and relative spacings in the ranges, say, $r=0.2$ to $0.9$ and
$s=0.3$ to $2.0$ (although in some cases up to $s=8$). For large
enough $m_1$ and $m_2$ and small enough $r$, four new distinct
temperatures (beyond  $T_{1c}$) were identified in plots of the
specific heats (per lattice site) of the coupled system; see
Figs.~\ref{fig1}-\ref{fig4}. In decreasing magnitude these were
\be
T_{1c}>T_{1max}>T_c(r,s)>T_{2max}>T_{2c},
\ee
where $T_{1max}$ and $T_{2max}$ locate rounded but (for $m_1,m_2\gg8$)
increasingly sharp maxima,  while $T_c(r,s)$ locates an overall or
bulk critical point where the specific heat must diverge
logarithmically. However, the amplitude of this logarithmic
singularity vanishes exponentially fast\cite{HAY} with increasing
$(1-r)m_1m_2/(m_1+m_2)$, as indicated in the text following
(\ref{dfrs}). As a consequence, the divergence soon becomes
invisible on graphical plots: see Figs.~\ref{fig2} and \ref{fig3}.
Finally, $T_{2c}$ represents the bulk Ising critical point for
interactions $J_2$; consequently, when $m_2 \to\infty$, the
lower-$T$ (or weaker) maxima obey $T_{2max}\to T_{2c}$ which simply
corresponds to the weaker, coupling strips become infinitely wide.

In the experiments\cite{PKMGn,PG,PKMGpr} a large two-dimensional
lattice of the liquid helium boxes, at edge-to-edge separation
$L_2$ (say, $=s L_1$) with $L_2=1\,\mu$m to $4\,\mu$m, was connected
and, thereby, coupled to a greater or lesser degree, via, in the
later experiments, a ``two-dimensional helium film of thickness
33 nm.'' This film corresponds, in the alternating-strip Ising model,
with the weak strips that connect and couple the strong strips; in
this way one might hope to identify an {\it effective} $J_2$ from
the superfluid transition of the film, at say, $T_s < T_\lambda$:
see below. In the earlier experiments,\cite{PKMGn,PG} the connection
of the boxes was achieved via channels of width\cite{PKMGn}
$1\,\mu$m and depth 19 nm (for $L_1=1\,\mu$m  boxes) and of
width\cite{PKMGn,PG} $2 \mu m$ and depth 10 nm
(for $L_1=2\,\mu$m boxes); in the Ising context, this set of
channels then constitutes the weak system.

Now {\bf proximity effects} appear dramatically in the Ising context
via the fact --- clear in Figs.~\ref{fig2}-\ref{fig4} and, especially,
in Fig.~\ref{fig5} --- that although an isolated and finite 1D strip
must always have its specific heat peak {\it below} the corresponding
bulk critical temperature,\cite{HAY} namely, for the weak strips,
$T_{2c}$, the lower-$T$ peaks (associated with the weak strips) are
always located {\it above} $T_{2c}$. For the parameters we have used,
these positive shifts amount to a few percent;  more precisely, the
fractional shift is close to $0.89/m_2$. Evidently, the shifts must
be attributed entirely to the fact that the weak strips ``feel,''
very directly, the ordering effects of the already well ordered
strong strips.

In the experiments on liquid helium, since all observed features are
close to $T_\lambda$, we follow Gasparini and coworkers and use the
temperature deviation variable
\be
\dot{t}= (T/ T_{\lambda})- 1< 0\quad\text{for $T< T_\lambda$}.  
\ee
Then Fig.~7 of Ref.~\onlinecite{PKMGpr}, exhibits essentially the
same proximity effect! Specifically, while the specific heat maximum
of an isolated helium film occurs at
$\dot{t}_{2max}^\infty\simeq-2.4\times10^{-3}$, the presence of
already superfluid boxes of size $L_1=2\,\mu$m spaced edge-to-edge
at $L_2 = 4\,\mu$m apart raises the maximum in the film's specific
heat to $\dot{t}_{2max}\simeq-1.4\times10^{-3}$. That amounts to a
positive proximity shift of $0.1$\% of $T_\lambda$. While this is
quite small, the precision of the experiments is so great that the
effect is beyond question.

Another aspect of the proximity (not investigated directly in the
Ising strip system) is evident in Fig.~8 of Ref.~\onlinecite{PKMGpr}.
This shows observations of the superfluid density, $\rho_s(T)$, for
an isolated helium film; this vanishes (discontinuously) above the
corresponding lambda point at $\dot{t}_c=-3.0 \times 10^{-3}$. On
the other hand, in the presence of the $2\,\mu$m boxes separated
by $4\,\mu$m the superfluid density of the connecting film is
significantly enhanced. Furthermore, the transition point itself
rises, by 0.12\%\ of $T_\lambda$, to  $\dot{t}_c=-1.8\times10^{-3}$.
Even more dramatic are the observations of $\rho_s(T)$ shown in
Fig.~16 of Ref.~\onlinecite{PKMGpr} (or Fig.~4 of
Ref.~\onlinecite{PG}): in the presence of $L_1=2\,\mu$m boxes spaced
closer at $L_2 =2\,\mu$m edge-to-edge, the transition point of the
film rises to $\dot{t}_c =-18 \times10^{-3}$, ``a full decade closer
to  $T_\lambda$'' as Perron {\it et al}.\cite{PKMGpr} comment.

Beyond the proximity effects discussed, we have studied within the
model of alternating Ising strips, the {\bf enhancements} of the
maxima caused by the {\bf coupling} between the strips. These effects
can be made evident by first noting that merely on the basis of
finite-size scaling the specific heats should display rounded maxima
near to but, for the upper or strong maxima, displaced below
$T_{1c}$ --- the bulk critical point of the 2D Ising model with
interactions $J_1$. To detect the effects of the coupling, therefore,
we have defined in (\ref{enhancement}) the net enhancement,
${\cal E}(T)$, by subtracting the expected (and
known\cite{HAY,HAYFisher}) rounded maximum of an isolated strip
(for given $m_1$). The definition (22) also includes deductions
related to the lower maxima associated with the weaker strips;
but these are of negligible magnitude in the vicinity of $T_{1c}$.

Then, as seen clearly in Figs.~\ref{fig7}-\ref{fig11}, there are
significant residual contributions, due to the coupling, that
increase or enhance the upper rounded maxima well above the pure
scaling contributions. Further numerical explorations (see
Figs.~\ref{fig9}-\ref{fig12}) then demonstrate that the overall
net enhancement can, at least approximately, be  decomposed in to
a finite background piece of order $1/(m_1+m_2)$ plus quite narrow
although rounded peaks near $T_{1c}$ and $T_{2c}$ of magnitude of
order $\ln(m_i)/(m_1+m_2)$ for $i=1,2$, respectively. The location
of the upper peaks is, in all cases, given roughly by
\be
t_{1max}\approx-0.893/m_1.
\label{t1shift}\ee
At this point these various conclusions, while in our view fully
convincing, lack support from exact asymptotic theory. Nevertheless,
it is certainly clear theoretically\cite{FisherFerd} that the
regularly spaced seams or grain boundaries along which the strong
and weak strips meet, must give rise to corrections asymptotically
of order at least $1/(m_1+m_2)$.

For the experiments on liquid helium the analogous enhancement
effects arising from the coupling are evident in Figs.~13 and 18
of Ref.~\onlinecite{PKMGpr} (and Fig.~3 of Ref.~\onlinecite{PG}).
Specifically, Fig.~13 for $L_1=1\,\mu$m  boxes coupled via
channels (of width $1\,\mu$m, depth 19 nm, with $L_2=1\,\mu$m)
shows a relatively narrow but well determined specific heat peak
needed to correct for the lack of scaling which is, otherwise,
expected for well isolated boxes of this size. Then, Fig.~18
shows an enhancement form of quite similar shape and magnitude
when $L_1=2\,\mu$m boxes at separation $L_2=2\,\mu$m are
coupled via a 33 nm film. The enhancement here, in fact, increases
the peak height by about 9\%\ (relative to uncoupled boxes) while
the peak location is again below $T_\lambda$ at approximately
$\dot{t}_{max}=-20\times 10^{-6}$. If this displacement is compared
with the Ising result (\ref{t1shift}) one might conclude that an
appropriate match would require $m_1$ of order $40,000$; this is
several times larger than the previous estimate of an appropriate
value of $m_1$ (in the third paragraph of this Section). This
difference might, however, be related mainly to the distinctly
different dimensionalities entailed in the helium and Ising systems;
that, in turn, along with the different dimensionality of the order
parameter, is an effect hard to guess.

Finally, however, it is clear that while the behavior of the
alternating layered Ising model reflects quite directly many of
the novel proximity and coupling features uncovered in the
striking experiments of Gasparini and coworkers for liquid
helium\cite{PKMGn,PG,PKMGpr} the quantitative features differ
considerably. More specifically, while the range of relative
separations, $s=m_2/m_1$, explored numerically compares well
with that relevant in the experiments (where, essentially,
$L_2/L_1=1$ or $2$), the strength ratio $r$, which in our study
has been confined to $r< 0.9$, should be much closer to unity
to match the experimental data. One might, for example, use the
observed values of the superfluidity onset temperatures relative
to $T_\lambda$ and derive an estimate for $r$ from the ratios
of $T_c(r, s)/T_{1c}$, etc. Similarly, one might regard the observed
maximum of the specific heat of an isolated  33 nm helium film as
providing an estimate of $T_{2c}$ in the model and hence of the
ratio $r=T_{2c}/T_{1c}=J_2/J_1$. Implementing these suggestions
leads to values of $(1-r)$ of order $3\times10^{-3}$. In this
regime of very small $(1-r )$, the separate rounded peaks
associated with $T_{1c}$ and $T_{2c}$ may, indeed, not be realized,
as already clear for $(1-r)=0.1$ in Fig.~\ref{fig2}. Clearly, the
experiments represent a rather different region of the underlying
parameter space than that which we have explored.

\begin{acknowledgments}
The authors thank F.M.\ Gasparini for extensive discussions and
correspondence, which stimulated this work. The help of J.H.H.\ Perk
on many tricky typesetting problems is gratefully acknowledged.
One of us (HA-Y) has been supported in part by the National Science
Foundation under grant No.\ PHY-07-58139.
\end{acknowledgments}

\end{document}